\documentclass[12pt]{article}
\usepackage{amsmath, bm}
\usepackage{amssymb, amsthm}
\usepackage{algorithm2e}
\usepackage{graphicx}
\usepackage{hyperref}
\usepackage{setspace}
\usepackage{subcaption}
\usepackage[numbers]{natbib}
\usepackage{url} 
\usepackage{enumitem}
\usepackage{xcolor}
\usepackage{authblk}
\usepackage{comment}
\usepackage{tabularx}
\usepackage{makecell}
\usepackage{bbm}
\usepackage{float}
\usepackage{rotating}
\usepackage{geometry}
\usepackage{pdflscape}
\newcommand{\blind}{0}

\begin{document}

\def\spacingset#1{\renewcommand{\baselinestretch}%
{#1}\small\normalsize} \spacingset{1}

\date{}
\if0\blind
{ 
  \title{\bf Quantifying patient and neighborhood risks for stillbirth and preterm birth in Philadelphia with a Bayesian spatial model}
  
\author{Cecilia Balocchi$^{1*}$, 
Ray Bai$^{2*}$, 
Jessica Liu$^{3}$,
Silvia P. Canel\'{o}n$^{4}$, Edward I. George$^{3}$, 
Yong Chen$^{4}$, and Mary R. Boland$^{5\#}$ \\
\small{$^{1}$School of Mathematics, University of Edinburgh, Edinburgh EH8 9YL, United Kingdom} \\
\small{$^{2}$Department of Statistics, University of South Carolina, Columbia, SC 29208, USA} \\
\small{$^{3}$Wharton School of Business, University of Pennsylvania, Philadelphia, PA 19104, USA} \\
\small{$^{4}$Perelman School of Medicine, University of Pennsylvania, Philadelphia, PA 19104, USA} \\
\small{$^{5}$Department of Mathematics, Saint Vincent College, Latrobe, PA 15650, USA} \\ \vspace{.5cm}
\small{$^{\#}$Corresponding author: Mary R. Boland, 300 Fraser Purchase Rd, Latrobe, PA 15650 (mary.boland@stvincent.edu)} \\
\small{$^{*}$Cecilia Balocchi (cecilia.balocchi@ac.ed.uk) and Ray Bai (rbai@mailbox.sc.edu) are joint first authors.}}

  \maketitle
} \fi

\if1\blind
{
  \begin{center}
    {\LARGE\bf Title}
\end{center}
} \fi

\begin{abstract}
Stillbirth and preterm birth are major public health challenges. Using a Bayesian spatial model, we quantified patient-specific and neighborhood risks of stillbirth and preterm birth in the city of Philadelphia. We linked birth data from electronic health records at Penn Medicine hospitals from 2010 to 2017 with census-tract-level data from the United States Census Bureau. We found that both patient-level characteristics (e.g. self-identified race/ethnicity) and neighborhood-level characteristics (e.g. violent crime) were significantly associated with patients' risk of stillbirth or preterm birth. Our neighborhood analysis found that higher-risk census tracts had 2.68 times the average risk of stillbirth and 2.01 times the average risk of preterm birth compared to lower-risk census tracts. Higher neighborhood rates of women in poverty or on public assistance were significantly associated with greater neighborhood risk for these outcomes, whereas higher neighborhood rates of college-educated women or women in the labor force were significantly associated with lower risk. Several of these neighborhood associations were missed by the patient-level analysis. These results suggest that neighborhood-level analyses of adverse pregnancy outcomes can reveal nuanced relationships and, thus, should be considered by epidemiologists. Our findings can potentially guide place-based public health interventions to reduce stillbirth and preterm birth rates. 
\vspace{.5cm}

\end{abstract}

\spacingset{1.5} 

\section*{Introduction}

Despite advancements in maternal health, adverse pregnancy outcomes such as stillbirth and preterm birth continue to be major public health problems. While there has been a 90\% reduction in infant mortality over the past century \citep{Mattison2001}, rates of stillbirth (i.e. fetal death at or after 20 weeks of pregnancy) have remained largely unchanged \citep{Kramer2003,Starling2020}.  For example, in 2013, a total of 23,595 stillbirths were reported in the United States (U.S.) \citep{Macdorman2015}. This represents a country-wide statistic and therefore the number of stillbirths in any individual city for a given year is often very sparse. Preterm birth (i.e. birth before 37 weeks of pregnancy) is another adverse pregnancy outcome that has gradually increased over time. 
The Centers for Disease Control and Prevention (CDC) estimated that in 2014, preterm birth affected one of every 10 infants in the U.S. \citep{Ferre2016}. The U.S. preterm birth rate also increased from 9.57\% in 2014 to 9.93\% in 2017 \citep{Martin2018}.

Stillbirth and preterm birth are significantly associated with increased patient risk of neonatal mortality and morbidity, adverse neuro-developmental and cognitive outcomes, and increased health care costs \citep{Muraskas2008,Xu2016}. Furthermore, the occurrence of adverse perinatal outcomes is often geographically heterogeneous, with some areas that are much more severely affected than others \citep{South2012, ZahriehOlesonRomitti2019a}. Thus, some researchers have argued that it is critical to quantify geographic regions of excess risk of adverse pregnancy outcomes in order to better develop targeted intervention strategies for subpopulations at highest risk \citep{South2012, ZahriehOlesonRomitti2019a}.

Recently, there has been growing interest in place-based interventions and policies for addressing health disparities \citep{McGowan2021, GaoAJE2023}. Place-based interventions are approaches for improving population health within a defined geographic location, delivered at a local or regional level rather than at a national level \citep{McGowan2021}. These intervention programs are often implemented at a neighborhood level \citep{Sharpe2013}. For example, city councils may designate specific neighborhoods as ``priority neighborhoods'' for targeted investment \citep{Sharpe2013}. In our study, we define a neighborhood as a spatial unit (e.g. a census tract) within a larger geographic area such as a city or a state. 

Our study is motivated by the growing interest in the public health community in place-based interventions to reduce stillbirth and preterm birth rates \citep{Nelin2023, CommonwealthFund2021, Project20, Khan2023}. To this end, we developed a Bayesian spatial model to separately analyze stillbirth and preterm birth and to quantify the patient-specific and neighborhood risk of both outcomes in the city of Philadelphia. Philadelphia is a particularly compelling case study because it is the sixth largest city in the United States and has been experiencing population growth and changes to its built environment for the first time in decades \citep{Balocchi2019}.  Our model incorporates information from adjacent neighborhoods in order to improve estimates for individual sites. Based on the estimated neighborhood probabilities of stillbirth and preterm birth, we stratified the Philadelphia census tracts into ``lower-risk,'' ``moderate-risk,'' and ``higher-risk'' neighborhoods for each outcome and analyzed the marginal associations between census-tract-level characteristics (e.g. racial composition, education levels, and crime incidences) and neighborhood risks.

One of the key findings of our analysis was that higher-risk neighborhoods had 2.68 times the average risk of stillbirth and 2.01 times the average risk of preterm birth compared to lower-risk neighborhoods. Our neighborhood analysis unveiled not only geographic patterns in stillbirth and preterm birth, but also the marginal neighborhood-level relationships that were missed by the patient-specific risk analysis. This is because the patient-specific risk analysis was based on conditional comparisons among different risk factors. In other words, the associations of each risk factor were determined conditionally on the other covariates in the model. As a result, high correlations between covariates might mask the true associations between some of the predictor variables and the outcomes of interest. For example, since poverty and public assistance levels are highly correlated, a joint regression analysis may determine that only one of these risk factors is significantly associated with preterm birth or stillbirth, conditionally on the other risk factor already being in the model. 
Contrastingly, our neighborhood risk analysis enabled us to identify higher-risk and lower-risk groups of neighborhoods and to compare the marginal risk factors between these different groups of neighborhoods. These neighborhood variations would not be ascertained from performing only a patient-specific risk analysis. This reinforces the need for epidemiologists to consider neighborhood-level risk analysis, in addition to patient-level risk analysis.

Our work builds upon and goes beyond recent work by \citet{YangAJE2023} who also quantified neighborhood risks of preterm birth in Philadelphia. However, \citet{YangAJE2023} only considered individual patient characteristics and did not include census-tract-level risk factors in their study. On the other hand, our study identified several neighborhood stressors significantly associated with the risk of both preterm birth and stillbirth, such as violent crime. Secondly, whereas \citet{YangAJE2023} only studied preterm birth, we also quantified stillbirth risk in our study. To our knowledge, our study is the first to quantify neighborhood risks of stillbirth in Philadelphia. While we focused on Philadelphia, our methodology can be applied to any city and can be used to guide place-based public health interventions to reduce stillbirth and preterm birth rates at a neighborhood level.

\section*{Methods} 

\subsection*{Patient data} 

The birth data from this study were obtained from hospitals within the University of Pennsylvania Health System (also known as Penn Medicine) from 2010 to 2017. The University of Pennsylvania Institutional Review Board approved this study with a waiver of informed consent (institutional IRB protocol number 828000). We used a previously developed and validated algorithm by \citet{Canelon2021} to identify deliveries from Penn Medicine electronic health records (EHRs). Within each EHR, the following delivery outcomes were annotated: stillbirth and preterm birth (each coded as either ``1'' or ``0''). In addition, the EHRs also reported each patient's residential address, the age of the patient at the time of delivery, binary variables for the patient's self-reported racial/ethnic group (Hispanic, non-Hispanic White, non-Hispanic Black or non-Hispanic Asian) and a binary variable for multiple birth (e.g. twins, triplets). In our analysis, we used White as the baseline category for the race/ethnic variables. 

After excluding patients with either missing values in the patient address or who lived outside of Philadelphia, we matched the remaining patients' residential addresses to their specific longitude and latitude coordinates. Based on these coordinates, we mapped each patient to one of the 384 U.S. census tracts in Philadelphia. Because some census tracts were very sparsely populated (e.g. the areas around rivers), we removed tracts that had fewer than 10 deliveries in each of the eight years. This was done mainly out of disclosure risk concerns, and it seems sensible because if a neighborhood only contained (for example) 4 deliveries, then having just 1 stillbirth in this neighborhood would greatly overestimate that neighborhood's risk of stillbirth.

\begin{table}[t]
		\caption{The 19 covariates that we used in our analysis of stillbirth and preterm birth in Philadelphia. \label{tab:covariates}}
	\spacingset{1}
	\centering 
	\resizebox{\textwidth}{!}{
		\begin{tabular}{lll}
			\hline
			\textbf{Patient-Level} & \textbf{Description}  \\ \hline
			age & Age of the patient  \\ 
			Black & Indicator if the patient self-identifies as Black  \\
			Hispanic & Indicator if the patient self-identifies as Hispanic  \\
			Asian & Indicator if the patient self-identifies as Asian  \\ 
			multiple birth & Indicator if multiple babies were delivered  \\
			\hline 
			\textbf{Neighborhood-Level} &  \textbf{Description}  \\
			\hline
			proportion Asian & Proportion of neighborhood that is Asian  \\ 
			proportion Hispanic & Proportion of neighborhood that is Hispanic  \\
			proportion Black & Proportion of neighborhood that is Black   \\
			proportion women & Proportion of women aged 15-50  \\ 
			poverty & Proportion of women aged 15-50 below the poverty level   \\
			public assistance & Proportion of women aged 15-50 who received public assistance  \\
			labor force & Proportion of women aged 16-50 who were in the labor force    \\
			recent birth & Proportion of women who gave birth in the past 12 months \\
			high school grad & Proportion of women aged 15-50 who graduated high school   \\ 
			college grad & Proportion of women aged 15-50 with a Bachelor's degree \\
			occupied housing & Total number of occupied housing units (log-transformed)   \\
			housing violation & Total number of housing violations (log-transformed)  \\
			violent crime & Total number of violent crimes (log-transformed)  \\
			nonviolent crime & Total number of nonviolent crimes (log-transformed)  \\ 
			\hline
	\end{tabular}}
\end{table}

\subsection*{Neighborhood data}

We downloaded data from the U.S. Census Bureau from 2010 through 2017 documenting racial makeup, poverty status, education level, and number of housing units for each census tract in Philadelphia. We also used data from the Philadelphia Police Department which reported the total number of violent and nonviolent crimes in each census tract from 2010 to 2017 \citep{Balocchi2019}. Web Appendix A gives the precise details of these data sources. Violent crimes included murders, rapes, robberies and aggravated assaults, whereas nonviolent crimes included burglaries, larceny, motor vehicle thefts and arson \citep{FBI2018}. Table \ref{tab:covariates} describes all 19 patient-level and neighborhood-level covariates included in our study. 

\subsection*{Statistical methods} 

Let $y^{q}_{ij}$ be a binary outcome with ``1'' indicating that patient $j$ in the $i$th neighborhood experienced stillbirth (for $q=1$) or preterm birth (for $q=2$). We assumed a mixed effects logistic regression model where $y^{q}_{ij} \sim \text{Bernoulli}(p^{q}_{ij})$ and
\begin{equation*} 
	\text{logit} \left( p^{q}_{ij} \right) = \alpha^{q}_i + \bm{x}_{ij}^\top \bm{\beta}^{q}.
\end{equation*}
Even though the model is fitted separately for each maternal health outcome, we subsequently drop the superscript $q$ for notational simplicity.
In this model $\bm{x}_{ij}$ is a vector that contains the values of the 19 covariates in Table \ref{tab:covariates} for patient $j$ in neighborhood $i$, $\bm{\beta}$ is a vector of unknown regression coefficients, and $\alpha_i$ is a neighborhood-specific random effect. Since we analyze the data in a Bayesian framework, we require prior distributions for the $\alpha_i$'s and $\bm{\beta}$. We used the conditional autoregressive (CAR) prior of \citet{Leroux2000} for the random effects. The CAR prior assumes the $\alpha_i$'s are spatially correlated and induces stronger spatial autocorrelation between geographically adjacent neighborhoods, thus facilitating principled sharing of neighborhood information \citep{Balocchi2019, Leroux2000}. The CAR prior contains some hyperparameters which we further endowed with weakly informative priors. We also endowed $\bm{\beta}$ with a weakly informative multivariate normal prior. Web Appendix B provides a detailed description of all the priors in our model.

We fitted our Bayesian spatial model separately for stillbirth and preterm birth as the outcome using a Markov chain Monte Carlo (MCMC) algorithm (described in Web Appendix D). For each outcome, we ran two MCMC chains of 5500 iterations each. We removed the first 500 iterations of each chain as burn-in and thinned the remaining samples by keeping only 1 of every 10 samples. This left us with a total of 1000 samples from the two chains. MCMC diagnostics reported in Web Appendix D confirm that this number of iterations was sufficient to achieve convergence and that our thinned samples were not too correlated.

\subsection*{Patient-specific risk analysis}

To conduct inference about patient risk factors for stillbirth and preterm birth, we used the posteriors for $\bm{\beta}$ from the mixed effects logistic regression model to compute the Bayesian probability of direction (PD)  \citep{makowski2019indices}. For the $k$th risk factor, the PD is the maximum of the posterior probability that the corresponding regression coefficient $\beta_k$ is less than zero or greater than zero. Thus, the PD is always between 0.5 and 1 and measures the certainty that $\beta_k$ is either positive or negative \citep{makowski2019indices}. A larger PD ($\geq 0.95$) indicates greater certainty that the $k$th risk factor is significantly associated with stillbirth or preterm birth, while a lower PD ($< 0.95$) indicates less certainty.

\subsection*{Neighborhood risk analysis}

Using the MCMC samples of $\alpha_i$ and $\bm{\beta}$, we first estimated the posterior distribution of the risk probability $p_{ij}$ for each $j$th patient living in the $i$th census tract as $p_{ij} = e^{\theta_{ij}} / (1+e^{\theta_{ij}})$, where $\theta_{ij} = \alpha_i + \bm{X}_{ij}^\top \bm{\beta}$ is the patient's log odds. We then obtained each $i$th neighborhood's risk probability $p_i$ by taking the mean of the $p_{ij}$'s for all patients living in the $i$th neighborhood. We approximated the posterior distributions for the $p_i$'s using the MCMC samples of the $p_{ij}$'s.  

We further stratified the Philadelphia census tracts in our study into three risk categories. Following \citet{Byrnes2015}, we used $k$-means clustering to cluster the  posterior mean neighborhood risks as ``lower-risk,'' ``moderate risk,'' or ``higher-risk.'' The cluster assignments were used to estimate the posterior distributions for the predicted probabilities of the lower-risk, moderate-risk, and higher-risk clusters, respectively denoted as $p_{LR}$, $p_{MR}$, and $p_{HR}$. These three cluster risk probabilities were determined by taking the mean of the $p_i$'s in each cluster. To assess marginal associations between neighborhood characteristics and neighborhood risk of stillbirth or preterm birth, we regressed the posterior mean cluster risk probabilities on the mean values for each neighborhood-level covariate within the three clusters.

\section*{Results} 

Starting with 63,334 patients, we removed 17,035 patients with either missing values in the patient address or who lived outside of Philadelphia. We further excluded 21 very sparsely populated census tracts with a total of 110 patients. Our final cohort contained 45,919 deliveries from 363 census tracts in Philadelphia, of which 385 (0.84\%) were stillbirths and 2897 (6.3\%) were preterm births. 

\subsection*{Model fitting and validation} 

We used the first six years of data from 2010 to 2016 to fit our Bayesian spatial model. The remaining year of data 2017 was used for model validation and for predicting the neighborhood risk probabilities. The reason that we estimated the neighborhood risk probabilities on an out-of-sample validation set was to avoid overfitting.

To validate the appropriateness of our method, we examined the posterior densities for the spatial autocorrelation parameter in the CAR prior. Web Appendix Figure \ref{fig:rhodensity} suggests that there was non-negligible spatial autocorrelation in the random effects, or that some unmeasured, spatially correlated variables might help explain the neighborhood variations in stillbirth and preterm birth. We also compared the model fit from our spatial model to a Bayesian mixed effects model with independent neighborhood random effects. These results (reported in Web Appendix C) also confirmed that our model with spatially correlated random effects provided a better fit to the data.

\subsection*{Results for patient-specific risk analysis}  

\begin{table}[t]
	\caption{Posterior mean odds ratios (OR) and
		the Bayesian probability of direction (PD) for stillbirth and
		preterm birth
		under the Bayesian spatial model.\label{tab:ORstillpre}}
	\spacingset{1}
	\centering
	\begin{tabular}{lrr|rr}
		& \multicolumn{2}{c|}{Stillbirth} & \multicolumn{2}{c}{Preterm birth} \\ 
		\hline
		& OR & PD & OR & PD  \\ 
		\multicolumn{1}{c}{\textbf{Patient-level}} & \multicolumn{2}{c|}{} & \multicolumn{2}{c}{}\\
		age & 1.01 & 0.79 & 1.00 & 0.95 \\ 
		Black & 2.23 & 1.00 & 1.55 & 1.00 \\ 
		Hispanic & 1.61 & 1.00 & 0.94 & 0.79 \\ 
		Asian & 0.97 & 0.58 & 1.12 & 0.84 \\ 
		multiple birth & 4.17 & 1.00 & 10.56 & 1.00 \\
		& & & & \\
		\multicolumn{1}{c}{\textbf{Neighborhood-level}}  & \multicolumn{2}{c|}{} & \multicolumn{2}{c}{}\\
		proportion Asian & 1.01 & 0.58 & 0.90 & 0.58 \\ 
		proportion Hispanic & 0.45 & 0.95 & 0.72 & 1.00 \\ 
		proportion Black & 0.61 & 0.95 & 0.92 & 0.74 \\ 
		proportion women & 0.24 & 0.89 & 1.88 & 1.00 \\ 
		poverty & 3.92 & 1.00 & 0.87 & 0.63 \\ 
		public assistance & 0.49 & 0.84 & 0.89 & 0.58 \\ 
		labor force & 1.25 & 0.58 & 0.49 & 1.00 \\ 
		recent birth & 1.40 & 0.74 & 1.48 & 0.79 \\ 
		high school grad & 1.16 & 0.58 & 2.63 & 1.00 \\ 
		college grad & 0.50 & 0.89 & 0.74 & 0.79 \\ 
		occupied housing & 0.99 & 0.53 & 0.93 & 0.68 \\ 
		housing violation & 0.88 & 0.84 & 0.99 & 0.68 \\ 
		violent crime & 1.59 & 1.00 & 1.16 & 1.00 \\ 
		nonviolent crime & 0.75 & 1.00 & 0.97 & 0.63 \\ 
		\hline
	\end{tabular} 
\end{table} 

\begin{figure}[t]
	\centering
	\includegraphics[width = 0.8\textwidth]{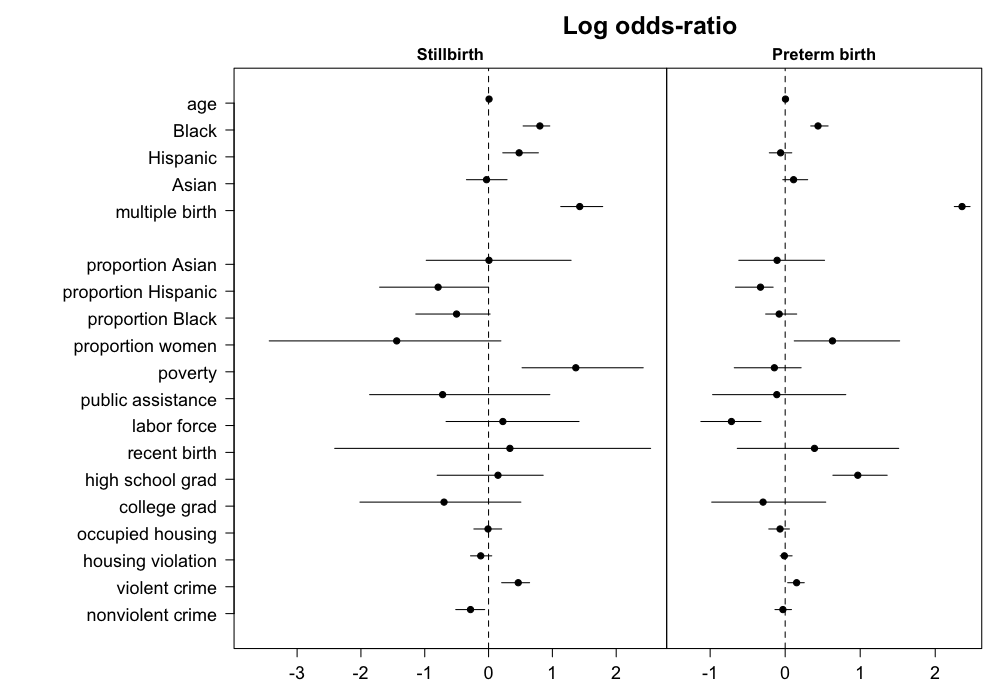}
	\caption{Posterior means and 95\% credible intervals of the log odd-ratios for the 19 covariates under our Bayesian spatial model. Left panel: results for stillbirth. Right panel: results for preterm birth. \label{fig:coeff_reweight}}
\end{figure}

Table \ref{tab:ORstillpre} reports the posterior mean odds ratio (OR) and the Bayesian PD for all 19 covariates. Figure \ref{fig:coeff_reweight} plots the 95\% posterior credible intervals of the log-odds ratio for these 19 covariates.  After controlling for other variables, patients who self-identified as Black were at much higher risk of stillbirth (OR=2.23, PD=1) and preterm birth (OR=1.55, PD=1), compared to White patients. Self-identified Hispanics were also at much higher risk of stillbirth (OR=1.61, PD=1) than White patients. Finally, patients experiencing multiple birth were at much higher risk of stillbirth (OR=4.17, PD=1) and preterm birth (OR=10.56, PD=1) than those giving birth to only one baby. These findings suggest that both maternal racial/ethnic group and multiple birth play an important role in determining a patient's likelihood of experiencing stillbirth or preterm birth.

Among the neighborhood-level risk factors, after controlling for other variables, an increase in neighborhood violent crime was significantly associated with greater risk of both stillbirth (OR=1.59, PD=1) and preterm birth (OR=1.16, PD=1). Increased neighborhood poverty also led to much higher patient-specific risk of stillbirth (OR=3.92, PD=1), while an increase in proportion of women who graduated from high school led to much higher patient-specific risk of preterm birth (OR=2.63, PD=1). Finally, an increase in the percentage of women in the labor force was associated with much lower patient-specific risk of preterm birth (OR=0.49, PD=1). 

\subsection*{Results for neighborhood risk analysis}

\begin{figure}
	\centering
	\includegraphics[width = 0.48\textwidth]{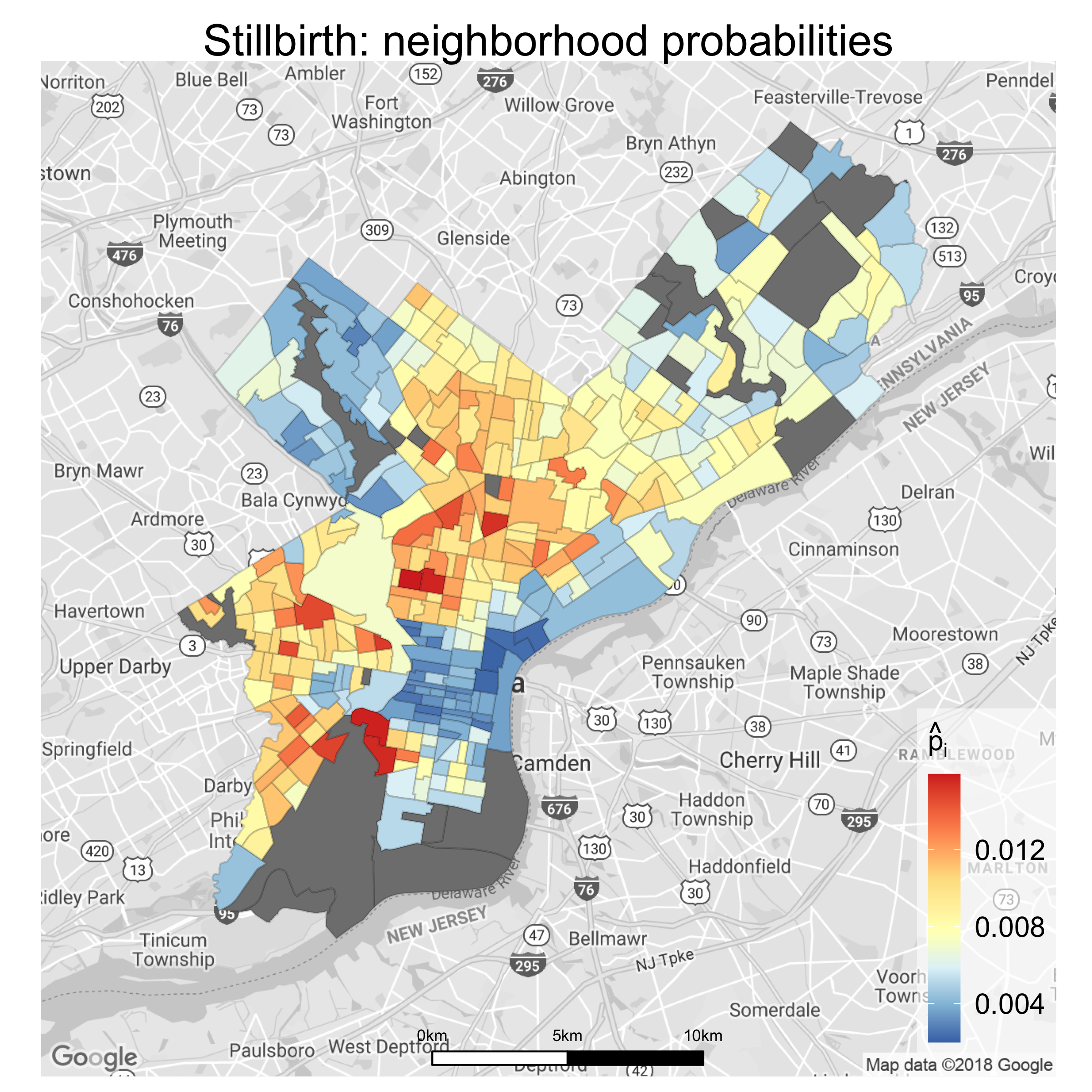}~ 
	\includegraphics[width = 0.48\textwidth]{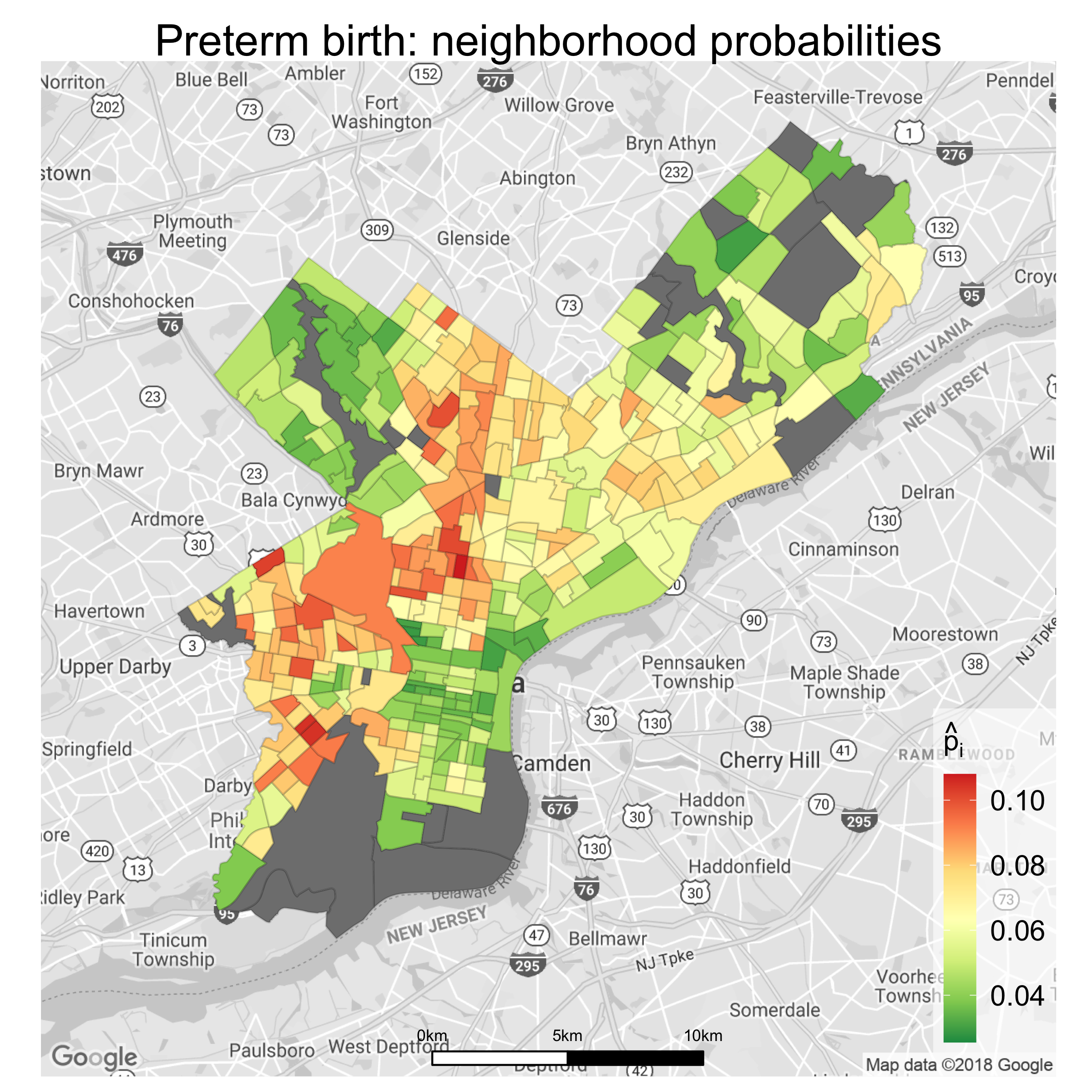} \\
	\includegraphics[width = 0.48\textwidth]{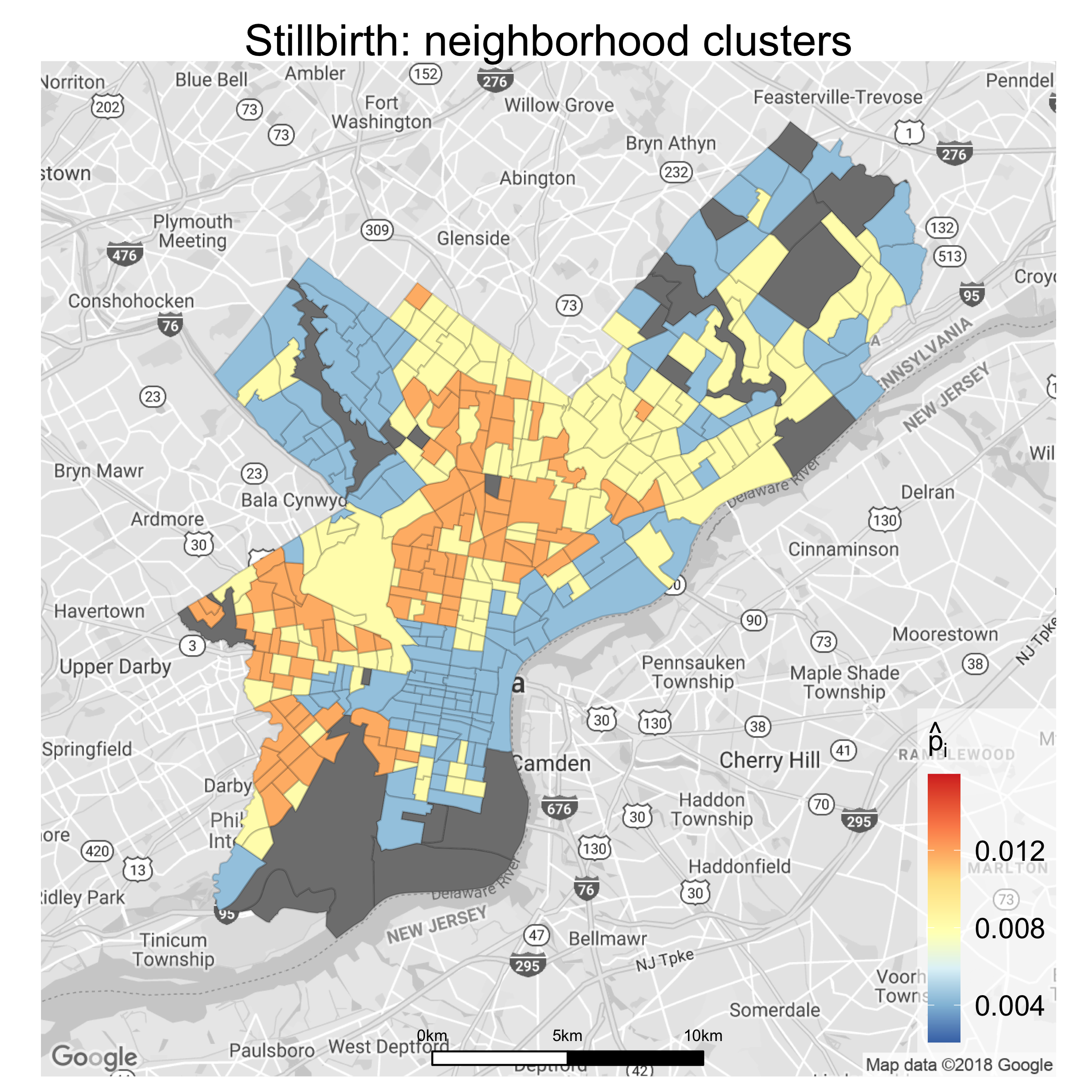}~ 
	\includegraphics[width = 0.48\textwidth]{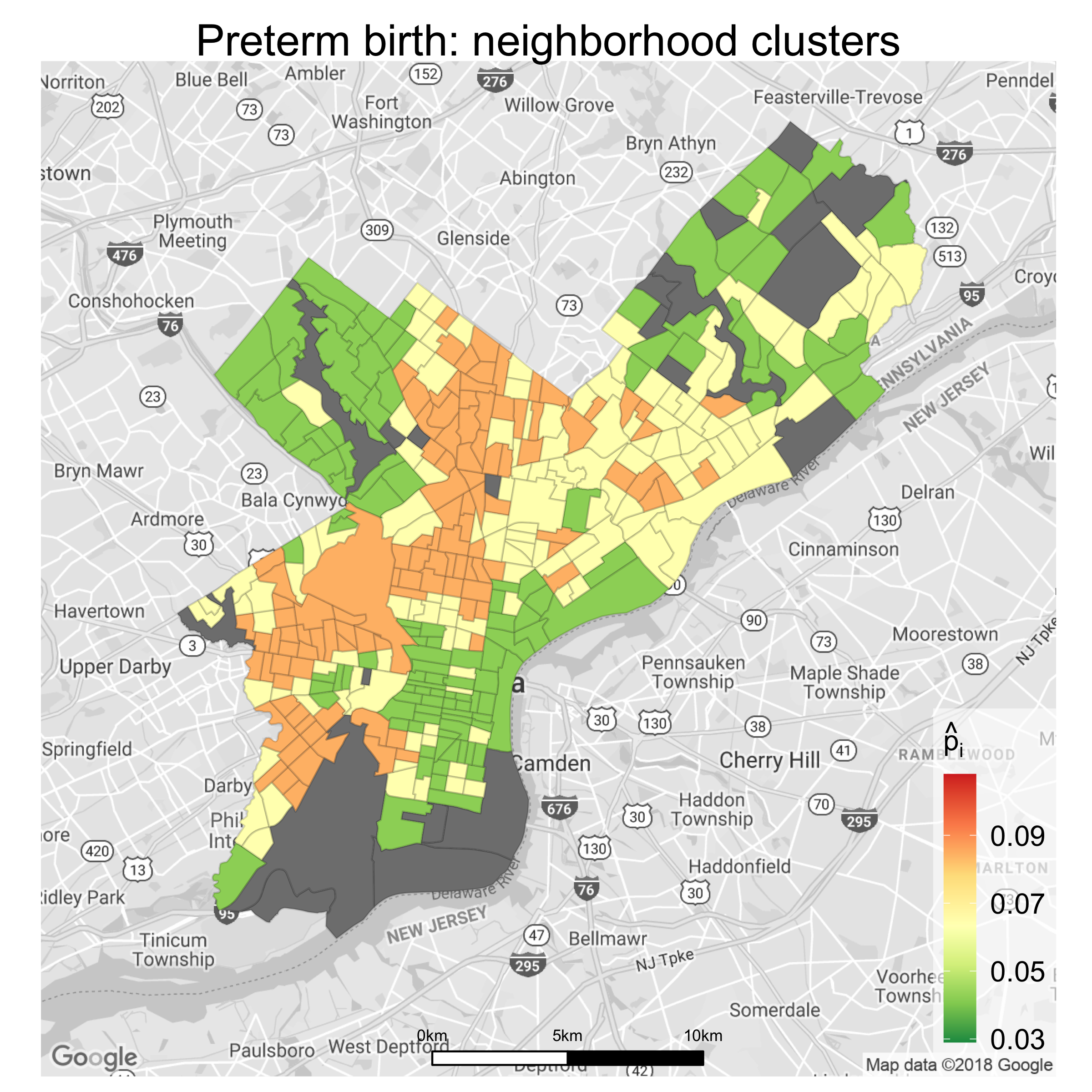}
	\caption{ {\bf Top two panels}:
		Maps of the predicted (posterior mean) neighborhood probabilities of having an adverse pregnancy outcome for stillbirth (left panel) and preterm birth (right panel). 
		{\bf Bottom two panels}:
		Maps of the clusters of lower-risk, moderate-risk and higher-risk neighborhoods  for stillbirth (left panel) and preterm birth (right panel). The orange clusters are the higher risk neighborhoods. The neighborhoods shown in gray are the very sparsely populated census tracts that were excluded from our analysis. 
		\label{fig:phat_clusters}
	}
\end{figure}

The top two panels of Figure \ref{fig:phat_clusters} plot the posterior means of the predicted neighborhood risk probabilities  for stillbirth and preterm birth. For stillbirth (top left panel of Figure \ref{fig:phat_clusters}), the census tracts with the highest risk of stillbirth were in North Philadelphia (such as North Philadelphia West), West Philadelphia (such as Cathedral Park, West Parkside, Mantua and Kingsessing) and South Philadelphia (such as Grays Ferry and West Passyunk). The neighborhoods with the lowest risk probabilities tended to be concentrated in Center City and areas immediately adjacent to Center City (such as Fishtown and Fairmount to the north and Bella Vista and Southwest Center City to the south) and in Northwest Philadelphia. Additionally, some small pockets of census tracts with lower risk were found in Northeast Philadelphia and West Philadelphia. 

The top right panel of Figure \ref{fig:phat_clusters} plots the posterior means of the predicted neighborhood probabilities for preterm birth. Similar to stillbirth, the census tracts with higher risk of preterm birth were concentrated in North Philadelphia (especially in North Philadelphia West, Upper North Philadelphia and East Germantown) and West Philadelphia (especially in Kingsessing, Cathedral Park and West Parkside). The neighborhoods with lower risk of preterm birth were concentrated in Center City and its surrounding neighborhoods and in Northwest Philadelphia, with some small pockets in Northeast Philadelphia and West Philadelphia.

The bottom two panels of Figure \ref{fig:phat_clusters} plot the cluster assignments for the 363 census tracts in our study and display a clear spatial pattern for the neighborhood risks of stillbirth and preterm birth in Philadelphia. Most of the higher-risk neighborhoods for stillbirth and preterm birth were concentrated in North Philadelphia and West Philadelphia. In addition, North Philadelphia and Northeast Philadelphia also contained many moderate-risk clusters. Table \ref{tab:cluster_cov} shows the number of neighborhoods in each risk category. 95 neighborhoods were categorized as ``higher-risk'' for stillbirth, and 112 neighborhoods were categorized as ``higher-risk'' for preterm birth.

\begin{figure}[t]
	\centering
	\includegraphics[width = 0.48\textwidth]{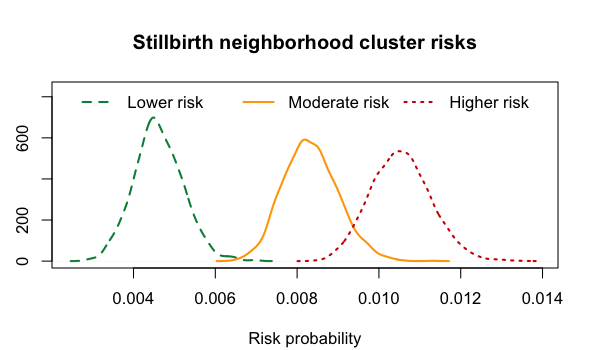}~ 
	\includegraphics[width = 0.48\textwidth]{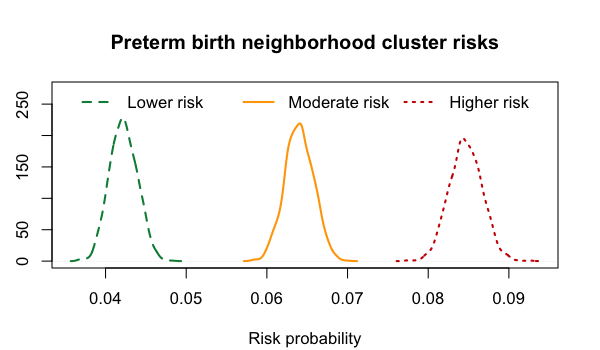}
	\caption{The neighborhood cluster risk posteriors for stillbirth ({\bf left panel}) and preterm birth ({\bf right panel}). The $95\%$ credible intervals for the cluster probabilities are respectively $(0.0034,0.0058)$, $(0.0071,0.0097)$ and $(0.0092,0.012)$ for stillbirth and $(0.039, 0.045)$, $(0.060,0.067)$ and $(0.081,0.088)$ for preterm birth.
		\label{fig:post_cl_phat}}
\end{figure}

Figure \ref{fig:post_cl_phat} plots the neighborhood cluster risk posterior distributions for $p_{LR}$, $p_{MR}$, and $p_{HR}$. The cluster risks for the higher-risk neighborhoods were on average more than double those for the lower-risk neighborhoods. In particular, the posterior mean risks for lower-risk neighborhoods were 0.0044 (95\% credible interval (CI): 0.0034, 0.0058) for stillbirth and 0.0422 (95\% CI: 0.039, 0.045) for preterm birth. These risks more than doubled to 0.0118 (95\% CI: 0.0092, 0.012) for stillbirth and 0.0847 (95\% CI: 0.081, 0.088) for preterm birth in the high-risk neighborhoods. Moreover, there was no overlap between the 95\% posterior credible intervals for $p_{LR}$ and  $p_{MR}$ for either stillbirth or preterm birth. This suggests that our neighborhood analysis was able to identify clusters with elevated (moderate or higher) average risk of stillbirth and preterm birth.

\begin{table}[t]
	\caption{Predicted (posterior mean) cluster risk probability, number of 
		neighborhoods in each cluster, and average neighborhood characteristics for the clusters of lower, 
		moderate and higher-risk neighborhoods. Plus and minus signs indicate the direction of each  
		of the neighborhood characteristics (i.e. whether they tend to increase or decrease with higher risk), 
		if they are determined to be significantly different from zero.\label{tab:cluster_cov}} 
	\spacingset{1}
	\centering
	\resizebox{\textwidth}{!}{
		\begin{tabular}{l | ccc | c | ccc | c}
			\hline
			& \multicolumn{4}{c|}{Stillbirth} & \multicolumn{4}{c}{Preterm birth} \\
			\hline
			& Lower  & Moderate & Higher  & Up/ & Lower  & Moderate & Higher  & Up/\\
			& risk   & risk     & risk    &  down  & risk   & risk     & risk    & down \\
			\hline
			Cluster risk & 0.44\% & 0.81\% & 1.18\% &  & 4.22\% & 6.40\% & 8.47\% &  \\ 
			Number of neighborhoods & 126 & 142 &  95 &  & 119 & 132 & 112 &  \\ 
			\hline
			proportion Asian & 0.09 & 0.07 & 0.03 & -- & 0.09 & 0.07 & 0.03 & -- \\ 
			proportion Hispanic & 0.07 & 0.14 & 0.14 & + & 0.08 & 0.18 & 0.08 &   \\ 
			proportion Black & 0.16 & 0.50 & 0.78 & + & 0.16 & 0.43 & 0.80 & + \\ 
			proportion women & 0.30 & 0.28 & 0.28 & -- & 0.29 & 0.28 & 0.29 &   \\ 
			poverty & 0.17 & 0.28 & 0.41 & + & 0.17 & 0.29 & 0.37 & + \\ 
			public assistance & 0.03 & 0.08 & 0.11 & + & 0.03 & 0.07 & 0.12 & + \\ 
			labor force & 0.77 & 0.68 & 0.63 & -- & 0.77 & 0.68 & 0.64 & -- \\ 
			recent birth & 0.04 & 0.05 & 0.06 & + & 0.04 & 0.06 & 0.06 & + \\ 
			high school grad & 0.17 & 0.30 & 0.35 & + & 0.16 & 0.30 & 0.35 & + \\ 
			college grad & 0.29 & 0.13 & 0.07 & -- & 0.31 & 0.13 & 0.07 & -- \\ 
			occupied housing & 7.26 & 7.28 & 7.31 &   & 7.31 & 7.29 & 7.24 &   \\ 
			housing violation & 3.87 & 4.45 & 5.00 & + & 3.75 & 4.51 & 4.93 & + \\ 
			violent crime & 3.91 & 4.54 & 5.08 & + & 3.81 & 4.62 & 4.97 & + \\ 
			nonviolent crime & 4.64 & 4.54 & 4.73 &   & 4.52 & 4.68 & 4.67 & + \\ 
			\hline
	\end{tabular}}
\end{table}

Table \ref{tab:cluster_cov} summarizes the main results from our marginal analysis of neighborhood risks. The ``Up/down'' column in Table \ref{tab:cluster_cov} indicates the sign of the slope of the regression if the slope was found to be significantly different from zero. More detailed results from this analysis are listed in Web Appendix Tables \ref{tab:cluster_orig_still} and \ref{tab:cluster_orig_pre}. Neighborhoods with higher risk of stillbirth and preterm birth tended to have higher proportions of Black residents, higher proportions of women living below the poverty line and on public assistance, and higher numbers of housing violations and violent crimes. Meanwhile, neighborhoods with lower risk of these outcomes tended to have higher proportions of Asian residents and higher proportions of women who were in the labor force or who had graduated from college. 

\section*{Discussion} 

We quantified both patient-specific and neighborhood risk of stillbirth and preterm birth in the city of Philadelphia. We applied a Bayesian spatial model to an EHR dataset of 45,919 deliveries at Penn Medicine hospitals from 2010 to 2017, augmented with spatially varying neighborhood data from the U.S. Census Bureau and the Philadelphia Police Department.   

Both patient-level and neighborhood-level risk factors were found to be significantly associated with patient-specific risk of stillbirth and preterm birth. Consistent with the literature, maternal racial/ethnic group and multiple birth were found to be significantly associated with increased risk of both of these outcomes \citep{YangAJE2023, ArechvoJCM2022, Sairam2002, FuchsSenat2016}. Previous research has linked neighborhood exposure to violent crime and small for gestational age infants, with the key finding being that across all racial/ethnic groups, there was a negative association between economic disadvantage and birth weight \citep{Masi2007}. Our findings support this prior work while building upon it. Namely, we found that neighborhood-level violent crime was also significantly associated with increased risk of two other important maternal health outcomes: stillbirth and preterm birth. 
Moreover, our results confirm previous research's findings on the relationship between poverty and higher risk of stillbirth \citep{jardine2021adverse}. These results could potentially guide public policies to reduce neighborhood stressors that may be triggering these adverse fetal outcomes.  

Going further than patient-level risk, we also provided measures of neighborhood risk assessment. We stratified census tracts into lower, moderate, and higher risk for stillbirth and preterm birth. We identified neighborhoods in West Philadelphia and North Philadelphia as being at higher risk for stillbirth and preterm birth. These findings were consistent with those of \citet{YangAJE2023} who had determined that these regions had greater risk of preterm birth. Earlier research had also found that residents in neighborhoods in North and West Philadelphia had lower average lifespans and higher rates of high cholesterol, hypertension, diabetes, and heart disease than Philadelphia as a whole \citep{ClosetoHome}. Our study adds to the literature on poor health outcomes in these same neighborhoods by also identifying them as having elevated risk of yet another outcome: stillbirth.

Our marginal neighborhood analysis revealed that higher-risk neighborhoods tended to have lower rates of women who had completed a college education or who were in the labor force. Higher-risk neighborhoods also tended to have higher rates of women who were living below the poverty line or who were on public assistance. Since we assessed the marginal comparisons of the risk factors in our neighborhood risk analysis, this seemingly contradictory result is explained by the collinearity between poverty and public assistance. When we inspect the patient-specific (i.e. conditional) analysis in Table \ref{tab:ORstillpre}, we find that conditional on poverty, public assistance does not have a significant association with the risk of stillbirth (PD=0.84) or preterm birth (PD=0.58).

These findings are important for two reasons. First, this demonstrates that our marginal neighborhood analysis can detect associations that are potentially missed by conditional patient-specific regression analyses due to multicollinearity. Secondly, previous work had found a protective effect of public assistance against COVID-19 in Philadelphia \citep{Boland2020}, whereas our findings suggest that public assistance may not have a similar protective effect against stillbirth or preterm birth. This underscores the importance of studying neighborhood-level factors and their contributions to specific health outcomes of interest, since the relationship between a neighborhood characteristic and disease risk may vary depending on the outcome. Our findings suggest that place-based policies to reduce poverty and violent crime may reduce neighborhood risk of stillbirth and preterm birth. Community-based programs to expand educational opportunities and labor force participation among women might also reduce neighborhood risks of stillbirth and preterm birth.

This study has several limitations. First, we treated each delivery as an individual patient, but there may have been patients who gave birth multiple times between 2010 and 2017. Second, our model used only residential addresses at the time of birth and did not account for previous home locations which may have affected the risk of stillbirth or preterm birth. 
Third, our estimated odds ratios may have been affected by some unmeasured confounders or measurement error in the neighborhood-level covariates, and we did not adjust for these. 
Fourth, we assumed only three risk clusters and used the deterministic $k$-means algorithm to determine cluster memberships. While our results suggested that the three neighborhood clusters were well-separated (Figure \ref{fig:post_cl_phat}), our analysis did not account for the uncertainty in either the estimated neighborhood risks or the number of clusters. Alternative techniques that do not require the practitioner to specify the number of clusters in advance, such as density-based spatial clustering with noise (DBSCAN) \citep{DBSCAN1996}, could be explored instead. Fifth, we did not consider potential temporal effects over time or spatio-temporal interactions, which may be able to better capture changes in stillbirth and preterm risk over time. This is the subject of ongoing work. 

In conclusion, we identified both patient-level and neighborhood-level risk factors for patient-specific risk of stillbirth and preterm birth. We further quantified neighborhood risks of these outcomes and identified associations between census-tract-level characteristics and risk of stillbirth and preterm birth that were not identified in the patient-level analysis.  Thus, neighborhood-level analysis should be considered by epidemiologists, in addition to patient-level analysis. This study is, to our knowledge, the first to quantify neighborhood risk of stillbirth at the census tract level in Philadelphia. Our findings have the potential to guide place-based interventions for reducing stillbirths and preterm births at a local neighborhood level.



 \section*{Funding}

 This work was funded partially by the European Research Council (Balocchi grant agreement No. 817257), the National Science Foundation (Bai DMS-2015528), the National Institutes of Health (Boland UL1-TR001878, Chen 1R01AI130460 and 1R01LM012607), and the Perelman School of Medicine (Boland).


\section*{Data availability}

The Penn Medicine EHR data contains individually identifiable information and may not be released. However, the Philadelphia neighborhood data was downloaded from \url{https://data.census.gov/cedsci/} and \url{https://opendataphilly.org} and is publicly available at \url{https://github.com/cecilia-balocchi/geospatial-adverse-pregnancy}. This GitHub repository also contains code to implement our Bayesian spatial model, along with simulated synthetic patient data, to run the code.

\section*{Acknowledgments}

This work was initiated when the first listed author was a PhD student at the Wharton School, University of Pennsylvania, under the mentorship of the fifth listed author and the second listed author was a postdoc at the Perelman School of Medicine, University of Pennsylvania, under the mentorship of the sixth and seventh listed authors. We also thank Phiwinhlanhla Ndebele-Ngwenya for her early work on a summer project (mentored by Dr. Boland) involving zip code level analysis that helped to motivate this expanded work.

\bibliographystyle{unsrtnat}
\bibliography{references_geographic_analysis}

\newpage

\section*{Web Appendix A: Additional data information} \label{Supp:S4}

While the patient-level data from Penn Medicine electronic health records (EHRs) is not publicly available, the neighborhood-level data were downloaded from \url{https://data.census.gov/cedsci/} in June 2020 and from \url{https://www.opendataphilly.org/dataset/crime-incidents} in January 2019. On \url{https://www.opendataphilly.org/}, the Philadelphia Police Department publicly releases the location, time, and type of each reported crime in the city. In Web Appendix Table~\ref{tab:datainfo}, we describe the precise data source for each neighborhood-level covariate in our dataset.
\vspace{1cm}

\begin{table}[H]
		\caption{Sources and data files for neighborhood-level covariates used in our analysis.
			ACS refers to the American Community Survey \label{tab:datainfo}}
	\centering
	\linespread{1} \small
		\resizebox{\textwidth}{!}{
		\begin{tabular}{lll}
			\hline
			Neighborhood-level covariate & Source & Data File \\
			\hline
			Proportion of each census tract that identifies as Asian Alone & ACS & B01001D \\
			Proportion of each census tract that identifies as Black or African-American & ACS & B01001B \\
			Proportion of each census tract that identifies as Hispanic or Latinx & ACS & B01001I \\
			Proportion of each census tract that identifies as White Alone & ACS & B01001A \\
			Proportion of women aged 15-50 years in each census tract & ACS & S1301 \\
			\makecell[l]{Proportion of women aged 15-50 years in each census tract below 100 percent \\ poverty level} & ACS & S1301 \\
			\makecell[l]{Proportion of women aged 15-50 years in each census tract that received \\ public assistance income in the past 12 months} & ACS & S1301 \\
			\makecell[l]{Proportion of women aged 16-50 years in each census tract that are \\ in the labor force} & ACS & S1301 \\
			\makecell[l]{ Proportion of women aged 15-50 years in each census tract who had a birth \\ in the past 12 months} & ACS & B13016 \\
			\makecell[l]{Proportion of women aged 15-50 years in each census tract that graduated  \\ high school (including equivalency)} & ACS & S1301 \\
			\makecell[l]{Proportion of women aged 15-50 years in each census tract that have a \\ Bachelor's degree} & ACS & S1301 \\
			Number of occupied housing units in each census tract & ACS & S2502 \\
			Housing Violations & OpenDataPhilly & \\
			Violent Crime Rate & OpenDataPhilly & \\
			Non-Violent Crime Rate & OpenDataPhilly & \\
			\hline
	\end{tabular}}
\end{table}

\newpage

\section*{Web Appendix B: Prior formulation for our method}

Recall that for patient $j$ in census tract $i$, we assumed a mixed effects logistic regression model,
\begin{equation*} 
	y_{ij} \sim \text{Bernoulli}(p_{ij}),~~ \text{logit} (p_{ij}) = \alpha_i + \bm{x}_{ij}^\top \bm{\beta},
\end{equation*}
where $\bm{x}_{ij}$ is a $p$-dimensional vector that contains the $p$ covariates for the $j$th patient in neighborhood $i$, $\bm{\beta} = (\beta_1, \dots, \beta_p)^\top \in \mathbb{R}^{p}$ is a vector of unknown regression coefficients (the fixed effects), and $\alpha_i$ is a random effect that accounts for the variation in neighborhood $i$ that cannot be explained by the $p$ covariates. 

We endowed the regression coefficients vector $\bm{\beta}$ with the weakly informative prior,
\begin{equation*} 
	\bm{\beta} \mid \bm{b}_0, \tau_{\beta} \sim \text{Normal} ( \bm{b}_0, \tau_{\beta}^2 \bm{I}_p ),~~\bm{b}_0 \sim \text{Normal} ( \bm{0}_p, 100 \cdot \bm{I}_p ),~~\tau_{\beta} \sim \text{Half-Cauchy} (0, 1).
\end{equation*} 
Let $\bm{\alpha} = (\alpha_1, \ldots, \alpha_n)^\top$ denote the vector of neighborhood-specific random effects in the mixed effects logistic regression model. In order to incorporate the neighborhood information in our spatial analysis, we employed a conditional autoregressive (CAR) prior on $\bm{\alpha}$ \cite{Besag1974, Lee2011, Leroux2000}. The CAR model is a Gaussian Markov random field which induces spatial dependence through an adjacency matrix for the areal units, which in our case study, are the census tracts of Philadelphia. We used the proper CAR formulation of Leroux \cite{Leroux2000}, which defines the distribution of each $\alpha_i$, given the other entries $\bm{\alpha}_{-i}$, as a normal distribution centered at a weighted average of a global mean $\alpha_0$ and the $\alpha_j$'s from neighborhoods that share a border with $\alpha_i$. That is, 
\begin{equation*}
	\alpha_i \mid \bm{\alpha}_{-i}, \alpha_0, \tau_{\alpha}, \rho \sim \text{Normal} \left( \frac{\rho \sum_j w_{ij} \alpha_j + (1-\rho) \alpha_0}{\rho \sum_j w_{ij} + (1-\rho)}, \frac{\tau_{\alpha}^2}{\rho \sum_j w_{ij} + (1-\rho)} \right),
\end{equation*}
where the $w_{ij}$'s are adjacency weights that are equal to one if the neighborhoods $i$ and $j$ share a border and equal to zero otherwise.  The autocorrelation parameter $\rho \in [0,1)$ represents the strength of spatial correlation between the components of $\bm{\alpha}$, with larger values of $\rho$ corresponding to stronger influence of bordering neighborhoods.

We collected the adjacency weights $w_{ij}$ into an adjacency matrix $\bm{W}$, constructed using shape files from the United States (U.S.) Census Bureau to determine which of the census tracts shared a border. The joint distribution of $\bm{\alpha}$ is then uniquely determined and can be written more compactly as
\begin{equation*}
	\bm{\alpha} \mid \alpha_0, \tau_{\alpha}, \rho \sim \text{Normal} \left( \alpha_0 \cdot \bm{1}, \tau_{\alpha}^2 \bm{\Sigma}_{\rm CAR} \right),
\end{equation*}
where $\bm{1}$ is a $n$-dimensional vector of all ones, $\bm{\Sigma}^{-1}_{\rm CAR} = \rho ( \bm{D}_{\bm{W}} - \bm{W} ) + (1-\rho) \bm{I}_n$, and $\bm{D}_{\bm{W}} - \bm{W}$ is the Laplacian matrix based on the neighborhood adjacency matrix $\bm{W}$ \cite{Balocchi2019}.
In our mixed effects logistic regression model, we used this prior for the random effects vector $\bm{\alpha}$. Note that if $\rho = 0$, then our model reduces to a GLMM with independent (non-spatially correlated) neighborhood random effects.  

In the CAR prior, there are three unknown parameters $(\rho, \alpha_0, \tau_{\alpha})$. We endowed the spatial autocorrelation parameter $\rho$ in the CAR prior with a standard uniform prior,
\begin{equation*}
	\rho \sim \text{Uniform}(0, 1).
\end{equation*}
The posterior distribution for $\rho$ can be used to assess the strength of spatial autocorrelation (see Web Appendix Figure \ref{fig:rhodensity}). We further endowed the grand mean $\alpha_0$ and the global scale parameter $\tau_{\alpha}$ in the CAR prior with weakly informative priors,
\begin{equation*} \label{CARhyperpriors}
	\alpha_0 \sim \text{Normal}(0, 100),~~\tau_{\alpha} \sim \text{Half-Cauchy}(0,1).
\end{equation*}
Our Bayesian model can be fit using Markov chain Monte Carlo (MCMC). See Web Appendix D.

\section*{Web Appendix C: Additional data analyses} \label{Supp:S1}

\subsection*{Assessing the appropriateness of a spatial model}

\begin{figure}[t]
	\centering
	\includegraphics[width = .8\textwidth]{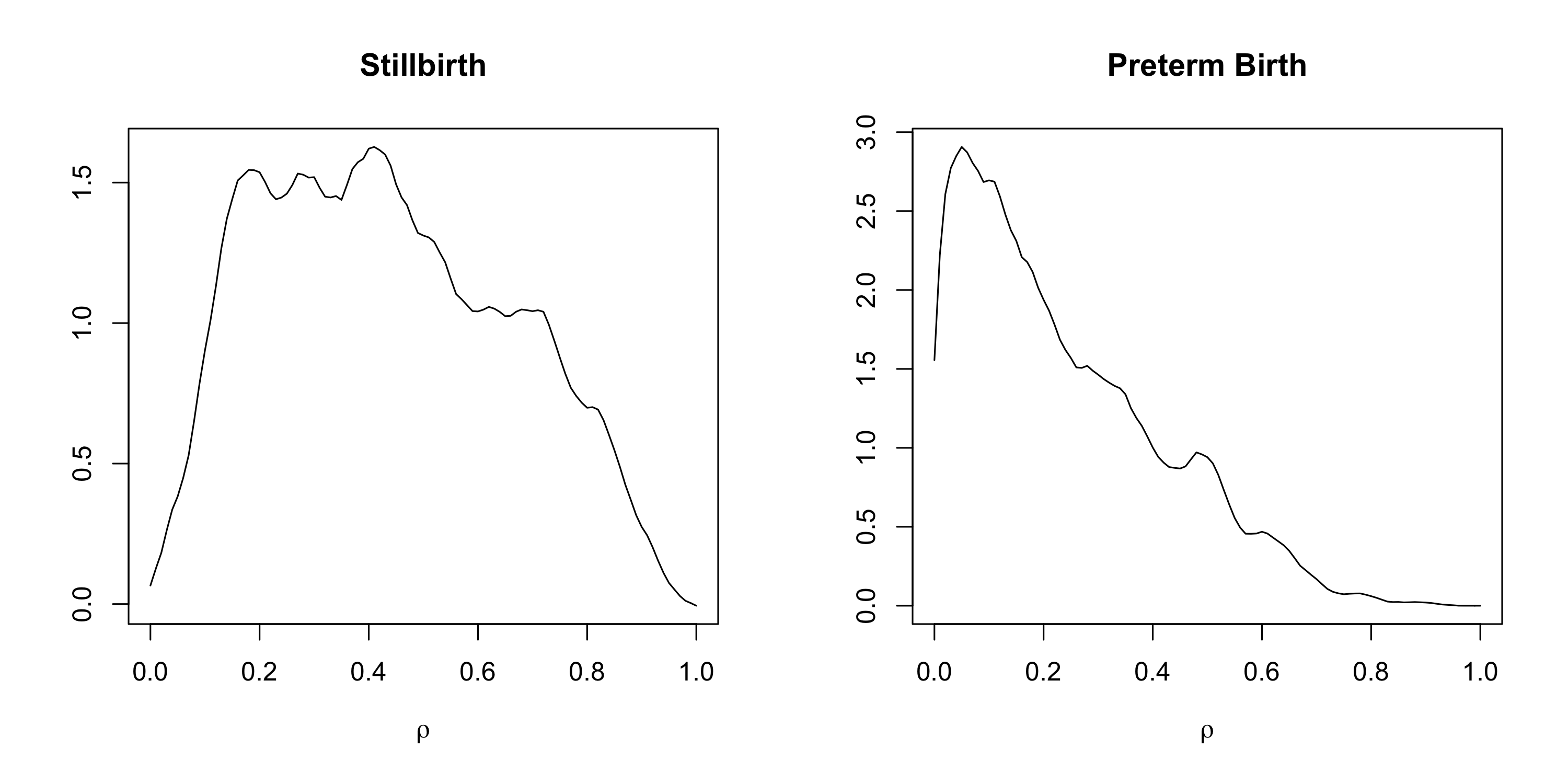}
	\caption{Plots of the posterior densities for autocorrelation parameter $\rho$ for stillbirth (\textbf{left panel}) and preterm birth (\textbf{right panel}) under the CAR model. \label{fig:rhodensity}}
\end{figure}

Web Appendix Figure \ref{fig:rhodensity} plots the posterior densities for the autocorrelation parameter $\rho$ under our proposed CAR method for stillbirth (left panel) and preterm birth (right panel). These posteriors were constructed using the MCMC samples of $\rho$ obtained from the MCMC algorithm described in Web Appendix D. The posterior for $\rho$ can be used to assess the strength of spatial autocorrelation in the data. 

For stillbirth, the left panel of Web Appendix Figure \ref{fig:rhodensity} shows that the majority of the posterior probability for $\rho$ was between 0.2 and 0.6, with non-negligible probability greater than 0.8. This suggests that the unexplained variation for stillbirth was spatially autocorrelated and that some unmeasured, spatially correlated covariates might help to explain the variation in stillbirth. On the other hand, for preterm birth, the right panel of Web Appendix Figure \ref{fig:rhodensity} shows that most of the posterior probability for $\rho$ was between 0 and 0.2, but there was still some posterior probability on values greater than 0.2. This suggests that there was weaker -- but possibly still non-negligible -- autocorrelation in the unexplained variation for preterm birth than stillbirth.

To further investigate the appropriateness of a CAR model for our data, we also compared the fit of our CAR model to a Bayesian model with independent random effects (i.e. $\rho$ fixed \textit{a priori} as $\rho=0$ in the CAR model). We compared these two models on the Philadelphia data using two separate Bayesian model selection criteria: the Deviance Information Criterion (DIC) \cite{Spiegelhalter2002} and the Watanabe-Akaike information criterion (WAIC) \cite{watanabe2010asymptotic}.

For an unknown parameter $\bm{\theta}$, the deviance is 
$D( \bm{\theta}) = -2 \log p (\bm{y} \mid \bm{\theta})$, where $p (\bm{y} \mid \bm{\theta})$ 
is the likelihood for the respective model. The DIC is given by $D(\bar{\bm{\theta}}) + 2 p_D$ where the first term is the deviance evaluated at the posterior mean of $\bm{\theta}$, 
and $p_D = \overline{D(\bm{\theta})} - D ( \bar{\bm{\theta}})$ is the effective number of model parameters where $\overline{D(\bm{\theta})} = E_{\bm{\theta} | \bm{y}} [ D (\bm{\theta})]$ is the posterior mean deviance. The DIC rewards better fitting models through the first term and penalizes more complex models through the second term.  The model with the smallest overall DIC value is preferred. 

As an alternative to DIC, one can also consider the WAIC. The WAIC is given by 
$\textrm{WAIC} = -2 \log \overline{ p (\bm{y} \mid \bm{\theta}) } + 2p_{W}$
where $\overline{ p (\bm{y} \mid \bm{\theta}) }$ is the posterior predictive density of the observed data, and 
$p_W = \sum_{i,j} \textrm{Var}[\log p(y_{ij} \mid \bm{\theta}) \mid \bm{y}]$, as recommended by \citet{gelman2014understanding}. Compared to DIC, WAIC considers the predictive density averaged over the posterior distribution, instead of conditioning on a point estimate, resulting in an approach more in line with the Bayesian framework.

\begin{table}[t]
		\caption{Comparisons of model fit for the CAR model vs. the independent random effects
			model using DIC and WAIC. \label{tab:model_comparison}}
	\centering
	\begin{tabular}{rcc|cc}
		&  \multicolumn{2}{c|}{Stillbirth} & \multicolumn{2}{c}{Preterm birth}\\
		\hline
		& CAR & Independent & CAR & Independent \\ 
		\hline
		DIC & 380.1 & 382.3 & 883.8 & 885.1 \\ 
		WAIC & 380.3 & 382.5 & 883.8 & 885.2 \\ 
		\hline
	\end{tabular}
\end{table} 

Let $\bm{\theta} = (\bm{\alpha}, \bm{\beta})$. For the mixed effects logistic regression model, the likelihood is defined as 
$p (\bm{y} \mid \bm{\theta}) = \prod_{i,j} p (y_{ij} \mid \bm{\theta}) = \prod_{i,j} \left[ p_{ij}(\bm{\theta})^{y_{ij}} (1-p_{ij}(\bm{\theta}))^{1-y_{ij}}\right],
$
where $p_{ij}(\bm{\theta}) = \exp(\alpha_i + \mathbf{x}_{ij}^\top \bm{\beta})/(\exp(\alpha_i + \mathbf{x}_{ij}^\top \bm{\beta})+1)$.
In practice, we estimate $\bar{\bm{\theta}}$, $\overline{D(\bm{\theta})}$, $\overline{ p (\bm{y} \mid \bm{\theta}) }$ and $\textrm{Var}[\log p(y_{ij} \mid \bm{\theta}) \mid \bm{y}]$ using MCMC samples from the algorithm described in Web Appendix D and then compute the DIC as $\textrm{DIC} = 2 \overline{D(\bm{\theta})} - D (\overline{\bm{\theta}})$ and the
WAIC as $\textrm{WAIC} = -2 \log \overline{ p (\bm{y} \mid \bm{\theta}) } +2 \sum_{i,j} \textrm{Var}[\log p(y_{ij} \mid \bm{\theta}) \mid \bm{y}]$. The model with the smallest DIC or WAIC provides the best model fit. 

We fit the CAR and the independent random effects to our training set and computed the DIC and WAIC using the validation set. The training and validation data are described in the Results section of the main article. Web Appendix Table \ref{tab:model_comparison} shows that the CAR model had a smaller DIC and WAIC for both stillbirth and preterm birth. This suggests that the CAR model is more appropriate to use than the independent random effects model for this particular dataset.

\subsection*{Additional results for the neighborhood risk analysis}

In the main article, we presented a comparative analysis of the neighborhood characteristics between ``lower-risk,'' ``moderate-risk,'' and ``higher-risk'' clusters of neighborhoods that we found. Here, we regressed the mean neighborhood covariate values of each cluster on the posterior mean predicted risk probabilities for each cluster. The main results of this analysis were reported in Table \ref{tab:cluster_cov} of the main manuscript. In Web Appendix Tables \ref{tab:cluster_orig_still} and \ref{tab:cluster_orig_pre}, we report more detailed results for stillbirth and preterm birth respectively. In the rightmost columns of Web Appendix Tables \ref{tab:cluster_orig_still} and \ref{tab:cluster_orig_pre}, we report the magnitudes of the estimated slope and the p-values from the simple linear regression models that we fit. 

In addition to these simple linear regressions, we also performed one-way analysis of variance (ANOVA) tests to test whether the mean neighborhood covariate values differed significantly across lower-risk, moderate-risk, and higher-risk groups of neighborhoods. We report the $F$-statistics and p-values from these ANOVA tests in the second rightmost columns of Web Appendix Tables \ref{tab:cluster_orig_still} and \ref{tab:cluster_orig_pre}. We see that there is agreement between the ANOVA tests and the linear regression models that we fit. Namely, if the ANOVA test determined that there was no significant difference between the lower-risk, moderate-risk, and higher-risk groups for a particular neighborhood characteristic (occupied housing and nonviolent crime for stillbirth; and proportion women, occupied housing, and housing violations for preterm birth), then the linear regression slope for that neighborhood characteristic was also \textit{not} significantly different from zero.
\vspace{1cm}

\begin{table}[H]
		\caption{Average neighborhood characteristics of the clusters of lower-risk, moderate-risk, 
			and higher-risk neighborhoods for \textbf{stillbirth}; results from three-way ANOVA tests and simple linear regression on the cluster risk probabilities.
			\label{tab:cluster_orig_still}}
	\centering
	\linespread{1} \small
	\resizebox{\textwidth}{!}{
		\begin{tabular}{l | c | c | c | cc | cc }
			\hline
			\textbf{Stillbirth} & Lower risk & Moderate risk& Higher risk &  \multicolumn{2}{c|}{ANOVA} & \multicolumn{2}{c}{Slopes}\\ \hline
			$\hat p$ & 0.44\% & 0.81\% & 1.18\% &  &  &  &  \\ 
			\hline
			& $\bar x_j$ & $\bar x_j$& $\bar x_j$ & F-value & p-value & $\beta$ & p-value  \\\hline
			proportion Asian &  0.0862 &  0.0673 &  0.0266 & 16.3005 &  0.0000 & -7.9247 &  0.0000 \\ 
			proportion Hispanic & 0.0729 & 0.1441 & 0.1367 & 6.3663 & 0.0019 & 9.2469 & 0.0042 \\ 
			proportion Black &   0.1606 &   0.5012 &   0.7776 & 144.7190 &   0.0000 &  84.1930 &   0.0000 \\ 
			proportion women &  0.2974 &  0.2808 &  0.2762 &  5.0093 &  0.0071 & -2.9566 &  0.0030 \\ 
			poverty &   0.1659 &   0.2828 &   0.4120 & 102.9180 &   0.0000 &  33.2907 &   0.0000 \\ 
			public assistance &  0.0328 &  0.0824 &  0.1130 & 58.5833 &  0.0000 & 11.0247 &  0.0000 \\ 
			labor force &   0.7675 &   0.6820 &   0.6320 &  48.3316 &   0.0000 & -18.6495 &   0.0000 \\ 
			recent birth &  0.0394 &  0.0546 &  0.0617 & 18.8219 &  0.0000 &  3.0859 &  0.0000 \\ 
			high school grad &   0.1690 &   0.3006 &   0.3521 & 110.1243 &   0.0000 &  25.4390 &   0.0000 \\ 
			college grad &   0.2945 &   0.1303 &   0.0699 & 195.1799 &   0.0000 & -31.2494 &   0.0000 \\ 
			occupied housing & 7.2564 & 7.2836 & 7.3087 & 0.4074 & 0.6657 & 7.1142 & 0.3668 \\ 
			housing violation &   3.8695 &   4.4477 &   5.0005 &  52.0092 &   0.0000 & 153.6266 &   0.0000 \\ 
			violent crime &   3.9053 &   4.5420 &   5.0788 &  95.5258 &   0.0000 & 159.9585 &   0.0000 \\ 
			nonviolent crime & 4.6378 & 4.5386 & 4.7254 & 3.0265 & 0.0497 & 9.7213 & 0.3644 \\ 
			\hline
	\end{tabular}}
\end{table}

\begin{table}[H]
		\caption{Average neighborhood characteristics of the clusters of lower-risk, moderate-risk, \\ and higher-risk neighborhoods for \textbf{preterm birth}; results from three-way ANOVA tests and \\ simple linear regression on the cluster risk probabilities. \label{tab:cluster_orig_pre}}
	\centering
	\linespread{1} \small
	\resizebox{\textwidth}{!}{
		\begin{tabular}{l | c | c | c | cc | cc }
			\hline
			\textbf{Preterm} & Lower risk & Moderate risk& Higher risk &  \multicolumn{2}{c|}{ANOVA} & \multicolumn{2}{c}{Slopes}\\ \hline
			$\hat p$ & 4.22\% & 6.40\% & 8.47\% &  &  &  &  \\ 
			\hline
			& $\bar x_j$ & $\bar x_j$& $\bar x_j$ & F-value & p-value & $\beta$ & p-value  \\\hline
			proportion Asian &  0.0870 &  0.0694 &  0.0306 & 15.8705 &  0.0000 & -1.3188 &  0.0000 \\ 
			proportion Hispanic &  0.0802 &  0.1794 &  0.0840 & 13.6561 &  0.0000 &  0.1865 &  0.7340 \\ 
			proportion Black &   0.1603 &   0.4263 &   0.8029 & 184.8069 &   0.0000 &  15.0807 &   0.0000 \\ 
			proportion women &  0.2904 &  0.2794 &  0.2870 &  1.3702 &  0.2554 & -0.0905 &  0.5923 \\ 
			poverty &  0.1671 &  0.2935 &  0.3712 & 67.5253 &  0.0000 &  4.8321 &  0.0000 \\ 
			public assistance &  0.0309 &  0.0749 &  0.1163 & 70.4655 &  0.0000 &  2.0140 &  0.0000 \\ 
			labor force &  0.7705 &  0.6797 &  0.6444 & 44.4952 &  0.0000 & -2.9971 &  0.0000 \\ 
			recent birth &  0.0412 &  0.0556 &  0.0567 & 10.8672 &  0.0000 &  0.3721 &  0.0000 \\ 
			high school grad &   0.1561 &   0.2999 &   0.3505 & 141.7264 &   0.0000 &   4.6243 &   0.0000 \\ 
			college grad &   0.3080 &   0.1338 &   0.0709 & 257.6153 &   0.0000 &  -5.6392 &   0.0000 \\ 
			occupied housing &  7.3053 &  7.2916 &  7.2418 &  0.6980 &  0.4982 & -1.4778 &  0.2668 \\ 
			housing violation &  3.7539 &  4.5071 &  4.9331 & 65.1284 &  0.0000 & 27.9321 &  0.0000 \\ 
			violent crime &   3.8085 &   4.6217 &   4.9665 & 108.7947 &   0.0000 &  27.5036 &   0.0000 \\ 
			nonviolent crime & 4.5159 & 4.6786 & 4.6676 & 2.9690 & 0.0526 & 3.6579 & 0.0428 \\ 
			\hline
	\end{tabular}}
\end{table}

\section*{Web Appendix D: MCMC algorithm and diagnostics} \label{Supp:S5}

\subsection*{MCMC algorithm} \label{Computation}

To fit our Bayesian CAR model, we sampled from the posterior distribution with MCMC. We used a Gibbs sampler to iteratively sample from the full conditional posteriors, together with the data augmentation strategy using P\'olya-Gamma latent variables, proposed by \citet{polson2013bayesian}. This strategy is the equivalent in logistic regression to the data augmentation strategy using latent Gaussian random variables in probit regression \citep{albert1993bayesian}.

Using the notation from \citet{polson2013bayesian}, we say that a random variable $X$ is distributed from a P\'olya-Gamma with parameters $b > 0$ and $c \in \mathcal{R}$ if 
$$
X \overset{D}{=} \frac{1}{2\pi^2} \sum_{k=1}^\infty \frac{g_k}{(k-1/2)^2+c^2/(4\pi^2)},
$$
where $g_k \sim \textrm{Ga}(b,1)$, and we denote $X \sim \text{Polya-Gamma}(b,c)$.

As shown in \citet{polson2013bayesian}, if $y_i \mid p_i \sim \text{Bernoulli}(p_i)$ and $\text{logit}(p_i) = \mathbf{x}_i^\top \bm{\beta}$ with $\bm{\beta} \sim \text{Normal}(\bm{b}, \bm{B})$, we can sample
\begin{align*}
	\omega_i \mid \bm{\beta} &\sim \text{Polya-Gamma}(1, \mathbf{x}_i^\top \bm{\beta}),\\
	\bm{\beta} \mid \bm{y}, \bm{\omega} &\sim \text{Normal}(\bm{m}_\omega,\bm{V}_\omega),
\end{align*}
with $\bm{V}_\omega = \left( \bm{X}^\top \bm{\Omega} \bm{X} + \bm{B}^{-1} \right)^{-1}$ and $\bm{m}_\omega = \bm{V}_\omega \left( \bm{X}^\top \bm{\Omega} \bm{\tilde y} + \bm{B}^{-1}\bm{b} \right)$, where $\bm{\Omega} = \text{diag}(\omega_i)$ and $\tilde y_i = (y_i-0.5) / \omega_i$.

To adapt this sampling scheme to our CAR model, we can write $\alpha_i + \mathbf{x}_{ij}^\top \bm{\beta} = \bm{z}_{ij}^\top \bm{\alpha} +  \mathbf{x}_{ij}^\top \bm{\beta}$, where $\bm{z}_{ij}$ is a $n$-dimensional vector with all entries equal to 0, except for the $j$th entry which is equal to 1. Then we can sample the parameters $\omega$, $\bm{\alpha}$, and $\bm{\beta}$ from their full conditionals,
\begin{align*}
	\omega_{ij} \mid \bm{\alpha}, \bm{\beta} &\sim \text{Polya-Gamma}(1, \alpha_i+ \mathbf{x}_{ij}^\top \bm{\beta}),\\
	\bm{\alpha} \mid \bm{y}, \bm{\beta}, \bm{\omega}, \tau_\alpha^2 &\sim \text{Normal}\left(
	\bm{V}_\alpha \left( \bm{Z}^\top \bm{\Omega} (\bm{\tilde y} - \bm{X} \bm{\beta} ) + \tau_\alpha^{-2} \bm{\Sigma}^{-1}_{\bm{\alpha}}\bm{1}\alpha_0\right), 
	\bm{V}_\alpha^{-1}
	\right),\\
	\bm{\beta} \mid \bm{y}, \bm{\alpha}, \bm{\omega}, \tau_\beta^2 &\sim \text{Normal}\left(
	\bm{V}_\beta \left( \bm{X}^\top \bm{\Omega} (\bm{\tilde y} - \bm{Z} \bm{\alpha} ) + \tau_\beta^{-2} \bm{b}_0 \right), 
	\bm{V}_\beta^{-1}
	\right),
\end{align*}
where $\bm{Z}$ is the $N \times n$ matrix of $\bm{z}_{ij}$, 
$\bm{V}_\alpha = \bm{Z}^\top \bm{\Omega} \bm{Z} + \tau_\alpha^{-2}\bm{\Sigma}^{-1}_{\bm{\alpha}}$, 
$\bm{V}_\beta = \bm{X}^\top \bm{\Omega} \bm{X} + \tau_\beta^{-2} \bm{I}$. Moreover, $\bm{\Sigma}^{-1}_{\bm{\alpha}}$ denotes the prior covariance matrix of $\bm{\alpha}$, which is equal to $\bm{\Sigma}^{-1}_{\rm CAR}$.

Under the CAR prior for $\bm{\alpha}$, we also need to sample the correlation parameter $\rho$. Note that this parameter affects the precision matrix of $\bm{\alpha}$, and its conditional distribution is given by 
$$
p(\rho \mid \text{rest}) = p(\rho \mid \bm{\alpha}, \alpha_0, \tau_\alpha^2) \propto \mid \bm{\Sigma}^{-1}_{\bm{\alpha}}(\rho) \mid^{1/2} \exp \left(-\frac{1}{2\tau_\alpha^2} (\bm{\alpha}-\alpha_0 \bm{1})^\top \bm{\Sigma}^{-1}_{\bm{\alpha}}(\rho) (\bm{\alpha}-\alpha_0 \bm{1})  \right).
$$
We sampled from this distribution using a Metropolis-Hastings (MH) within Gibbs step, with proposal density $g(\rho^* \mid \rho_{t}) = {\rm Beta}(\xi \cdot \rho_{t}/(1-\rho_{t}), \xi)$. This ensures that the mean is equal to $\rho_t$, and we choose $\xi=5$ so that $g(\rho^{\star} \mid \rho_t$) has small variance.

Assuming a half-Cauchy prior on $\tau_\alpha$ and on $\tau_\beta$, i.e. $p(\tau_\alpha) \propto (\tau_\alpha^2+s_\alpha^2)^{-1}$ and $p(\tau_\beta) \propto (\tau_\beta^2+s_\beta^2)^{-1}$, their conditional posterior distributions can be written as follows:
\begin{align*}
	p(\tau_\alpha \mid \bm{\alpha}, \alpha_0) &\propto (\tau_\alpha^2+s_\alpha^2)^{-1} \tau_\alpha^n \exp \left(-\frac{1}{\tau_\alpha^2} (\bm{\alpha}-\alpha_0 \bm{1})^\top \bm{\Sigma}_{\alpha}^{-1} (\bm{\alpha}-\alpha_0 \bm{1})/2 \right), \\
	p(\tau_\beta \mid \bm{\beta}, \beta_0) &\propto (\tau_\beta^2+s_\beta^2)^{-1} \tau_\beta^p \exp \left(-\frac{1}{\tau_\beta^2} \sum_{k=1}^p (\beta_k - b_{0,k})^2/2 \right).
\end{align*}
To sample from these posterior distributions, we used an MH step with a carefully designed proposal distribution. In general, given a random variable distributed according to $p(x) \propto (x^2 + s^2)^{-1} x^{-2\alpha}\exp(-\beta/x^2)$, consider the following proposal distribution, $q(x) = x^{-2\alpha-1}\exp(-\beta/x^2)$; note that $x \sim q$ is equivalent to $x^2 \sim \text{Inverse-Gamma}(\alpha, \beta)$. Moreover, if $\alpha$ and $\beta$ are carefully chosen, the acceptance ratio for the MH steps simplifies considerably:
$$
a(x \to \tilde x) = \frac{(x^2 + s^2) }{(\tilde x^2 + s^2) } \frac{\tilde x}{x}. 
$$
Thus, to sample $\tau_\alpha$ from its posterior, we simply need to generate from an inverse gamma distribution with parameters $\alpha = n/2$ and $\beta = (\bm{\alpha}-\alpha_0 \bm{1})^\top \bm{\Sigma}_{\alpha}^{-1} (\bm{\alpha}-\alpha_0 \bm{1})/2$, consider its square root as proposed value and compute the acceptance $a(\tau_\alpha \to \tilde \tau_\alpha)$. Similarly for $\tau_\beta$, we followed the same steps using $\alpha = p/2$ and $\beta = \sum_{k=1}^p (\beta_k - b_{0,k})^2/2$.

Finally, the full conditional for the remaining parameters are standard and can be easily sampled as follows:
\begin{align*}
	\alpha_0 \mid \bm{\alpha}, \tau_\alpha^2 &\sim \text{Normal}\left( \frac{\sum_{i=1}^{n} \alpha_i / \tau_\alpha^2}{1/ \tau_\alpha^2 + 1/100}, \frac{1}{1/ \tau_\alpha^2 + 1/100} \right), \\
	\bm{b}_0 \mid \bm{\beta}, \tau_\beta^2 &\sim \text{Normal}\left( \frac{\bm{\beta} / \tau_\beta^2}{1/ \tau_\beta^2 + 1/100}, \frac{1}{1/ \tau_\beta^2 + 1/100} \bm{I}_p \right). \\
\end{align*}

\subsection*{MCMC diagnostics} \label{app:newresults:MCMC}

Here, we report the diagnostics for the MCMC algorithm from fitting the CAR model to the Philadelphia dataset.  Web Appendix Table~\ref{tab:MCMC_diagnostics} shows the effective sample size (ESS), posterior mean and Monte Carlo standard error (MCSE) for the regression coefficients (i.e. log-odds ratio) of the individual and neighborhood covariates. Results are reported in Web Appendix Table~\ref{tab:MCMC_diagnostics} for stillbirth and preterm birth. 

\begin{figure}[H]
	\centering
	\includegraphics[width =\textwidth]{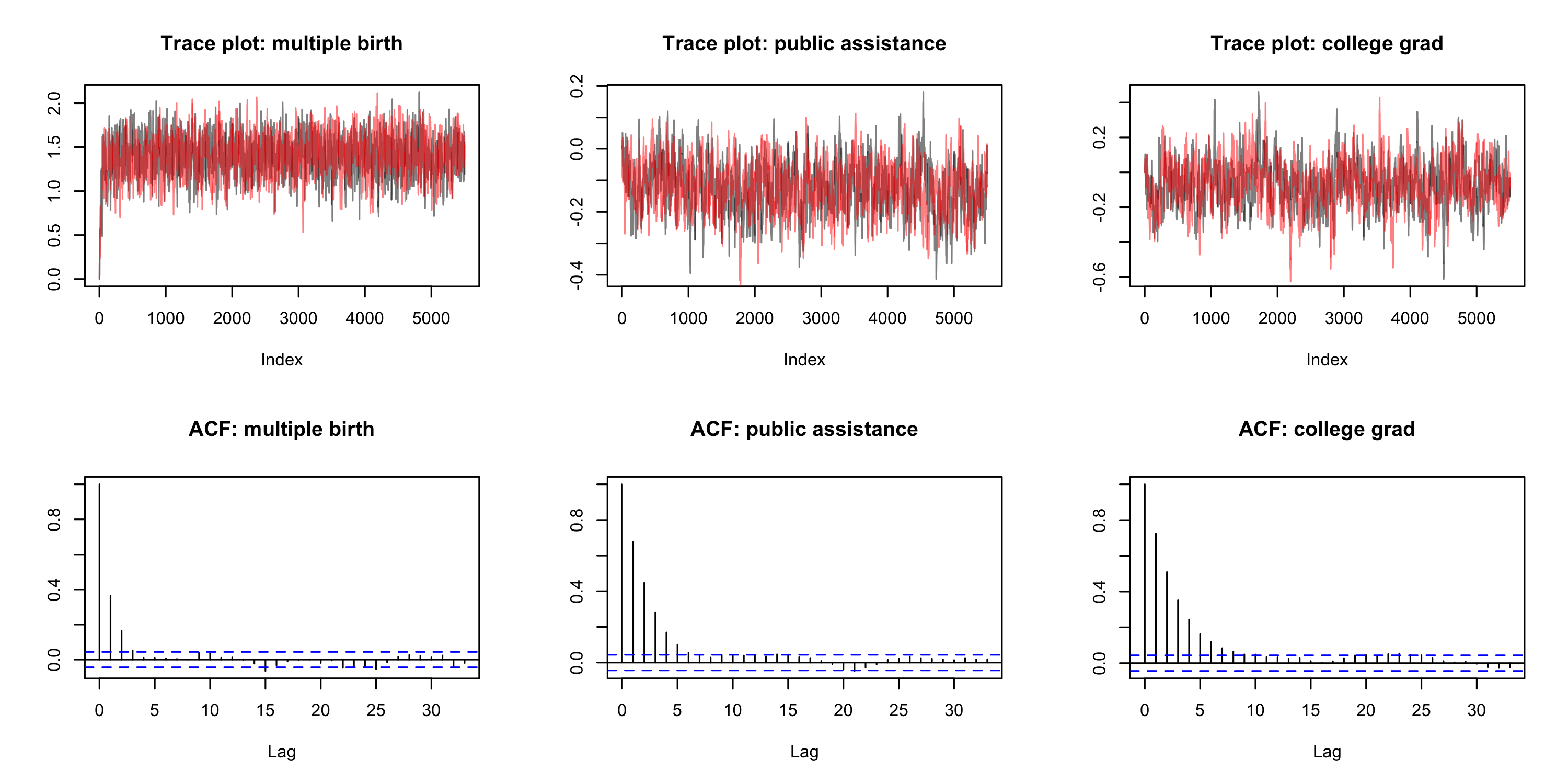} \\
	\includegraphics[width =\textwidth]{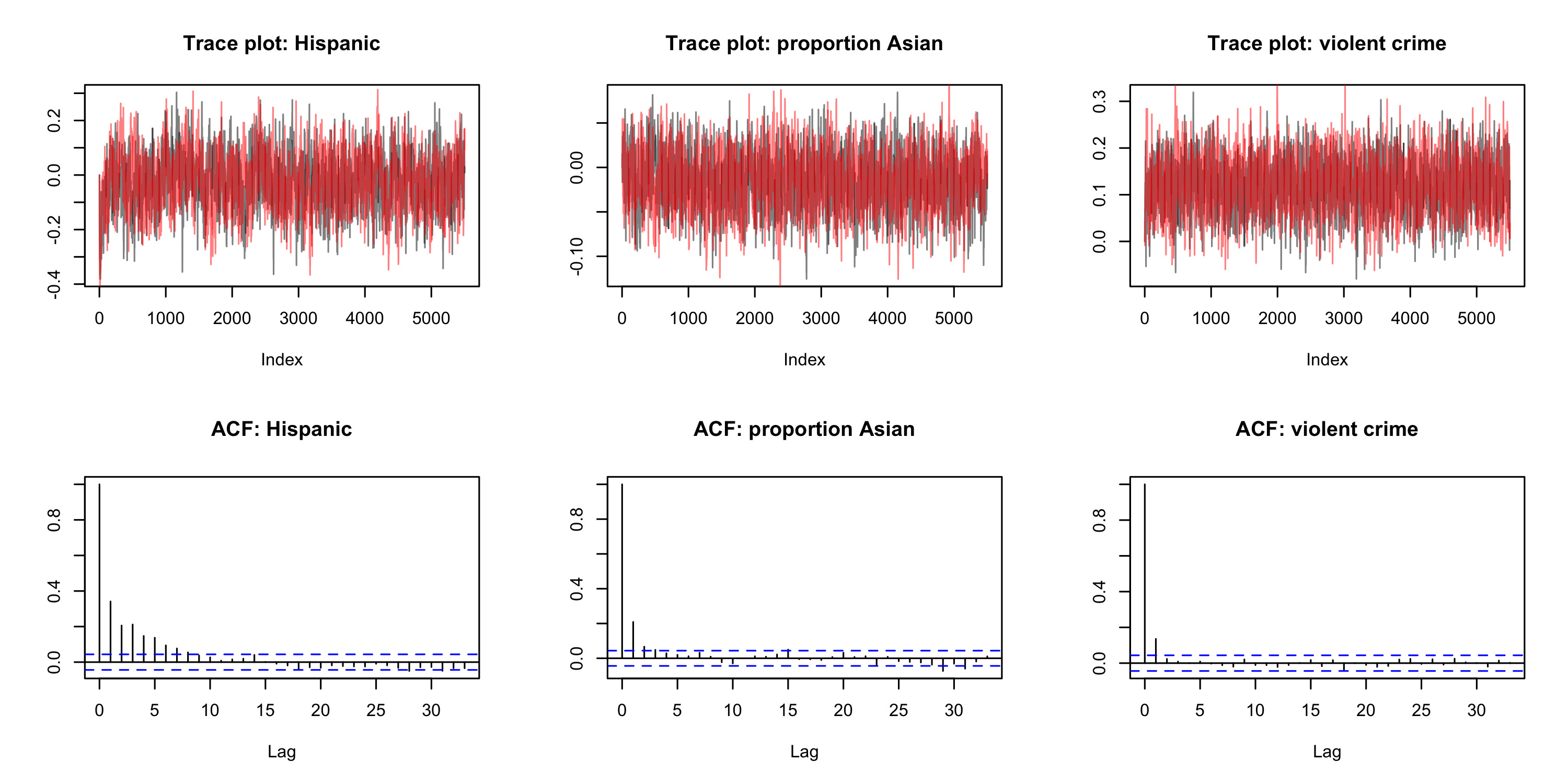}
	\caption{Trace plots of the two MCMC chains and autocorrelation plots of the thinned samples for several regression coefficients in our CAR model. In the top two rows, we plot the results for stillbirth, and in the bottom two rows, we plot the results for preterm birth. \label{fig:MCMCdiagnostics}} 
\end{figure}

Web Appendix Figure~\ref{fig:MCMCdiagnostics} shows the trace plots and autocorrelation plots of the MCMC samples for several of the regression coefficients under our CAR model. Web Appendix Figure~\ref{fig:MCMCdiagnostics} shows that the MCMC algorithm converged within 5500 iterations for the Philadelphia dataset. Moreover, by thinning the chain every 10 iterations, our final MCMC samples were not overly correlated.

\begin{table}[H]
		\caption{Effective sample size (ESS), posterior mean regression coefficient ($\beta_j$), 
			and Monte Carlo standard error (MCSE) for $\beta_j$ under the CAR model. 
			These
			diagnostics are reported for both stillbirth and preterm birth. \label{tab:MCMC_diagnostics}}
	\centering
	\begin{tabular}{r|rrr|rrr}
		\hline
		& \multicolumn{3}{c|}{Stillbirth} & \multicolumn{3}{c}{Preterm birth}\\
		\hline
		& ESS & $\beta_j$ & MCSE & ESS & $\beta_j$ & MCSE \\
		\hline
		Hispanic & 168.2 & -0.024 & 0.0239 & 662.0 & 0.449 & 0.0040 \\ 
		Black & 68.3 & 0.450 & 0.0238 & 231.9 & 0.630 & 0.0060 \\ 
		Asian & 190.3 & 0.131 & 0.0342 & 1011.8 & -0.354 & 0.0047 \\ 
		multiple birth & 943.7 & 2.356 & 0.0100 & 4241.4 & 1.408 & 0.0014 \\ 
		age & 468.4 & 0.020 & 0.0035 & 2511.8 & 0.029 & 0.0006 \\ 
		proportion Asian & 326.4 & -0.016 & 0.0060 & 1609.8 & -0.026 & 0.0011 \\ 
		proportion Hispanic & 379.6 & -0.053 & 0.0056 & 1839.4 & -0.068 & 0.0011 \\ 
		proportion Black & 293.8 & -0.035 & 0.0111 & 1183.4 & -0.061 & 0.0020 \\ 
		proportion women & 386.6 & 0.038 & 0.0050 & 2064.8 & -0.105 & 0.0008 \\ 
		poverty & 450.3 & -0.017 & 0.0059 & 2462.0 & 0.243 & 0.0010 \\ 
		public assistance & 430.3 & -0.003 & 0.0050 & 2249.7 & -0.124 & 0.0008 \\ 
		labor force & 399.5 & -0.059 & 0.0062 & 2042.1 & -0.050 & 0.0010 \\ 
		recent birth & 461.4 & 0.017 & 0.0037 & 2230.2 & 0.038 & 0.0006 \\ 
		high school grad & 432.0 & 0.109 & 0.0062 & 2212.4 & 0.068 & 0.0010 \\ 
		college grad & 356.7 & -0.074 & 0.0090 & 1780.7 & -0.068 & 0.0015 \\ 
		occupied housing & 377.2 & -0.015 & 0.0054 & 1821.4 & 0.063 & 0.0009 \\ 
		housing violation & 416.0 & -0.033 & 0.0058 & 1869.5 & -0.099 & 0.0011 \\ 
		violent crime & 413.7 & 0.120 & 0.0097 & 1881.7 & 0.166 & 0.0017 \\ 
		nonviolent crime & 375.0 & -0.027 & 0.0064 & 1836.8 & -0.086 & 0.0011 \\ 
		\hline
	\end{tabular}
\end{table}

\end{document}